\documentclass[twocolumn,pra, amsmath, aps,showpacs]{revtex4}
\newcommand{\bea}{\begin{eqnarray} }
\newcommand{\eea}{\end{eqnarray}}
\newcommand{\bean}{\begin{eqnarray*}}
\newcommand{\eean}{\end{eqnarray*}}
\newcommand{\nn}{\nonumber \\}
\def\bfg#1{{\mbox{\boldmath $#1$}}}

\def\od#1,#2{\frac{d#1}{d#2}}
\def\odz#1,#2{\frac{d^2#1}{d{#2}^2}}
\def\pd#1,#2{\frac{\partial #1}{\partial #2}}
\def\pdz#1,#2{\frac{\partial^2 #1}{\partial {#2}^2}}
\def\pdd#1,#2{\frac{\partial^3 #1}{\partial {#2}^3}}
\def\pdv#1,#2{\frac{\partial^4 #1}{\partial {#2}^4}}
\def\pdzz#1,#2,#3{\frac{\partial^2 #1}{\partial {#2}\partial{#3}}}

\def\eq#1{Eq.~(\ref{#1})}

\def\eqn#1{(\ref{#1})}

\def\B{{\bf B}}

\def\b{{\bf b}}

\def\e{{\bf e}}

\def\E{{\bf E}}

\def\g{{\bfg g}}

\def\v{{\bf v}}

\def\btimes{~{\bf \times}~}
\def\bnabla{{\bf \nabla}}
\def\bcdot{~{\bf \cdot}~}
\newcommand{\lbs}{\left (}
\newcommand{\rbs}{\right )}

\def\od#1,#2{\frac{d#1}{d#2}}
\def\hod#1,#2{\frac{\hat d#1}{d#2}}
\def\odz#1,#2{\frac{d^2#1}{d{#2}^2}}
\def\odd#1,#2{\frac{d^3#1}{d{#2}^3}}
\def\pd#1,#2{\frac{\partial #1}{\partial #2}}
\def\pdz#1,#2{\frac{\partial^2 #1}{\partial {#2}^2}}
\def\pdd#1,#2{\frac{\partial^3 #1}{\partial {#2}^3}}
\def\pdv#1,#2{\frac{\partial^4 #1}{\partial {#2}^4}}
\def\pdzz#1,#2,#3{\frac{\partial^2 #1}{\partial {#2}\partial{#3}}}

\usepackage{graphicx}
\usepackage{dcolumn}
\usepackage{bm}
\usepackage{epsfig}

\begin{document}


\bibliographystyle{unsrt}
%
%
\title{Rotating mirror with all-directional pinch compressions
\\
--- The  beauty and simplicity in controlled nuclear fusion
}
%
%
%
\author{Linjin Zheng} 

\affiliation{Institute for Fusion Studies, 
University of Texas at Austin, Austin, TX 78712}

\date{\today}

\begin{abstract}

Controlling nuclear fusion is so challenging that for decades, people have been asking: Are we closer to infinite clean energy? 
Upon the philosophy of beauty and simplicity,  the current work points out that there can be a shortcut: 
 the rotating mirror with detached electrodes and all-directional pinch compressions.
 This is based on the provisional patents filed recently by the University of Texas at Austin. 
The device combines the steady-state and fast processes in the two main streams of controlled
nuclear fusion research: magnetic confinement fusion and inertial confinement fusion. 
The fuel plasma is preheated in a steady-state process in a rotating mirror with detached electrodes
and then the pinch compressions in both radial and longitudinal directions are applied as the fast process. Preheating and longitudinal compression, in addition to the radial compression, significantly boost the nuclear fusion rate. Fast compression after the preheating limits the time for thermalization between ions and electrons and, therefore, minimizes the impact of electron 
bremsstrahlung radiation loss on ions.
Based on the existing experimental results,  the current method can be extrapolated to have the potential to reach or exceed the Lawson criterion for the first demonstration of the feasibility of peaceful usage of nuclear fusion energy.

\end{abstract}


\maketitle

\section{Introduction: The law above the natural laws}

There is an old saying that ``fusion is 30 years away, and always will be". 
The current paper elucidates how the beauty and simplicity considerations matter
and how the rotating mirror with detached electrodes and all-directional pinch compressions
can be a shortcut for controlling nuclear fusion as an energy source, 
based on the provisional patents filed recently by the University of Texas at Austin \cite{pt1,pt2}.

One of the famous quotes by A. Einstein is ``the most incomprehensible thing about the world is that it is at all comprehensible". Indeed, the vast and ever-changing universe is even governed by the natural laws. 
Our scientific endeavors become meaningful only because there are natural laws.
However, more awesome is that there is the Law above the laws of nature.

\begin{figure}[htp]
\centering
\includegraphics[width=80mm]{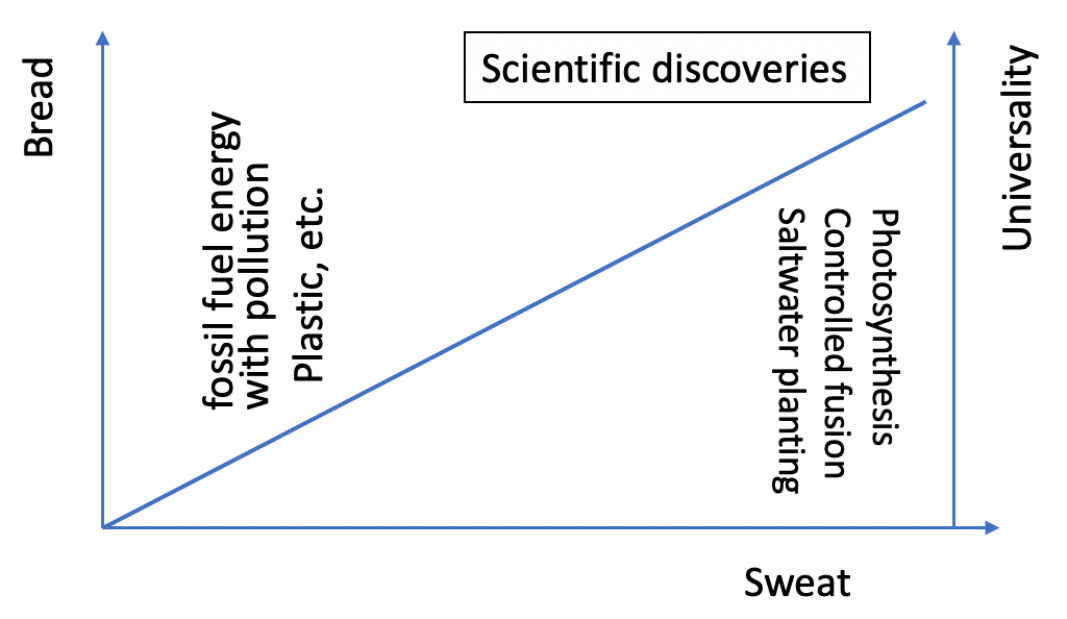}\vspace*{-2mm}
\caption{Distribution of scientific discoveries: 
the ``bread" versus ``sweat".}
\label{bread}
\end{figure}

The most serious challenges humans are facing are food shortages and diminishing energy resources. If natural photosynthesis could be reproduced, the issue of food shortages would be addressed. If controlled nuclear fusion could be achieved with a net energy yield, the energy resource problem would be solved. Unfortunately, the difficulties of achieving these goals are disproportionately large compared to the discoveries that are less critically important or that can be harmful to the natural environment. This reminds us of Genesis 3:19:
``In the sweat of thy face shalt thou eat bread", as depicted in Fig. \ref{bread}.
If the ``universality" is used as the second vertical axis,
 one can see that the processes are reversed. The hardest discoveries are often most popular naturally, while the easiest discoveries, like plastics, are often rare in nature \cite{adv,mhd0}. There is the Law governs the laws.

To solve the deadlock, to cross over the hill, the mountain, the inscription in the tower of the University of Texas at Austin reminds us `ye should know the truth and the truth shall make you free' (John 8:32). There is a saying: who tied the bell should be the one to untie it.  These lead us to humbly return to the Creator's wisdom as claimed in Einstein's another quote: “I want to know
God’s thoughts; the rest are details”. 
When the Universe was created, He claimed ‘it
is good’ and ‘it is so’ (Genesis 1). Beauty (the ‘good’)
and simplicity (the ‘so’) are embedded in nature. 

Looking up at the vast starry sky, the admiration, the endless yearning for the beauty of the universe may arise out of our hearts. However, think about that, if the speed of light were not constant but followed the Galileo transformation, the light from the stars moving away from us would be slowed down and the light from the stars moving toward us would be sped up.  The lights, when reaching our eyes from the splendid galaxies after travelling millions of light years, would be distorted. It is because the speed of light is constant that 
we can enjoy the beauty of the sky \cite{iopkinetic} and one has Einstein's theories of relativity.

The challenges of controlling nuclear fusion lie in its complexity. To fuse the nuclei,
fusion fuel must be heated to extreme temperatures exceeding 100 million degrees Celsius in order
to overcome the Coulumn repulsive force. The fuel gas becomes the plasma. 
No solid matter can hold the hot plasmas. The complicated magnetic field configuration
has to be constructed to confine it. What is more, the confinement is usually unstable. 
To deal with these difficulties, one has to think
about the Creator's wisdom: beauty and simplicity. Simplicity is regarded as the ultimate beauty.

In controlled nuclear fusion research \cite{fusion}, there are two main approaches: the magnetic confinement fusion approach with the steady-state process and the inertial confinement fusion approach with the fast process,
such as ITER \cite{iter} and the National Ignition Facility (NIF) \cite{nif}, respectively.
In the magnetic confinement fusion approach, a plasma containing fusion fuels is confined by a strong magnetic field and heated to thermonuclear fusion conditions. In the inertial confinement fusion approach, a target pellet of fusion fuel is compressed to thermonuclear fusion conditions by powerful laser or particle beams. The beams vaporize the outer layer of the shell and induce an implosion. The reaction force from the implosion presses the shell inwards and compresses the fuel for the fusion reaction.

Since neither the steady-state nor the fast process on its own is foreseen to easily 
reach the Lawson criterion \cite{lawson},
which states that the energy produced by fusion reactions must be greater than the energy lost to
the environment.
Combining the steady-state and fast processes for controlling nuclear fusion has been a dream to overtake the Lawson criterion.  Magnetic target fusion is an effort in this direction \cite{mtf}. One of the examples is as follows. The fuel plasma is preheated in a compact toroid and then compressed by an imploding metal liner using the conventional fluid mechanics principle. 

Combining the steady-state and fast processes is certainly a shortcut. However, using the magnetic field
for both processes makes the idea even simpler. 
This led to the concepts to shoot two preheated plasmoids by the pinch guns \cite{p4,p5,compfrc}.
The current work describes the concept of the rotating mirror with all-directional pinch compressions \cite{pt1,pt2}, especially, the longitudinal compression is facilitated 
by the non-spatially-static moving pinches. In contrast to
the concepts of shooting two preheated plasmoids, the current concept keeps the preheated
plasma in the central core while pinching it from all directions. It is even simpler with
an additional longitudinal compression factor in order of magnitude added. The conceptual novelty and engineering simplicity are features of the new concept.

The remaining arrangements of the paper are as follows:
In Sec. \ref{smirror}, the mirror concept is introduced in terms of the conceptual
 beauty and simplicity; In Sec. \ref{srot},  the rotating mirror 
with detached electrodes is described;  In Sec. \ref{spinch},
the rotating mirror with all-directional pinch compressions is presented;
The conclusions and discussion
will be given in Sec. \ref{scon}.

\section{The mirror concept, the beauty and simplicity}

\label{smirror}

In magnetic confinement fusion, there are three major magnetic confinement concepts: tokamak, stellarator, and mirror. The beauty and simplicity of the concepts determine their feasibility or achievability.
The confinement of hot plasma by the magnetic field is merely a force-balanced equilibrium. 
It can be unstable. It is like placing a ball on a hilltop. A small perturbation can cause
the ball to roll off the hill. This is the so-called instability.  From the Hamiltonian theory, 
one knows that the symmetry can lead to a motion invariant, which imposes
a constraint for instabilities to develop. 

Historically, the stellarator \cite{stell} was invented even earlier than the tokamak \cite{tokamak}, or about the same time. But the tokamak was successful first. Today, we know that the original stellarator design has not addressed the particle orbit loss issue. Although the early setbacks of stellarator experiments were due to the limitations in the theoretical understanding and computational capabilities of stellarator design at the time, they did make us think about why the tokamak concept was more likely to succeed,
under the same conditions and constraints.
The symmetry property of the tokamak concept, the ``beauty" feature, matters.

Both tokamak and stellarator concepts are based on toroidal confinement. They do not have the end loss 
issue as compared to linear confinement, such as a mirror. However, the toroidal confinement needs 
to generate the field line rotation transform to neutralize the charge separation due to the curvature drifts
of charged particles. Tokamak uses the induced toroidal current while stellarator
relies on twisting the coils.
Apparently, a stellarator has the advantage of not requiring a toroidal plasma current. However, it breaks the symmetry. This makes the existence of nested magnetic surfaces unguaranteed.
The current designs, e.g., W7-X \cite{w7x} or various new quasihelically symmetric configurations \cite{boozer}, have resorted to making the safety factor a ratio of large rational numbers, for example, $q \approx 100/99$. This only avoids the large islands by excluding their resonance surfaces. The magnetic shear becomes nearly zero in this approach. Consequently, the magnetic shear stabilization is absent. The other issue is that the current calculations are based on the guiding center description. A full orbit calculation often shows that the energetic particles cannot be confined. Breaking the symmetry causes the absence of conservation laws,
which affects the particle orbit confinement as well. Besides, a stellarator has an intrinsic equilibrium beta limit due to the Shafranov shift. 
In addition, stellarators are difficult to build from an engineering perspective.

The tokamak concept is well-known, such as ITER \cite{iter}. A tokamak needs a toroidal current to generate the field line rotational transform, and is therefore hard to maintain a steady-state confinement for a long duration. The Ohmic toroidal current also raises the system energy, leading to explosive instabilities, such as disruptions. Kink and tearing instabilities can also prevail due to the Ohmic current. Although it maintains the toroidal
symmetry, it is still more complicated than a mirror, a linear device.

The advantage of a mirror lies in its symmetry and simplicity \cite{post}. However, it has a bad curvature and a loss cone in the velocity space. Nevertheless, recent experiments, for example, in WHAM, confirmed that the plasma rotation can play a substantial stabilization role \cite{wham}. Theoretically, it is proved that
the rotation frequency in a mirror is usually not constant along the magnetic field lines \cite{equilibrium}
and the non-constant rotation can stabilize the interchange or flute modes \cite{stability}.
There are also other developments in the mirror research, including
the invention of the detached electrodes for driving the rotation  \cite{pt1}. 
All of these factors, as will be further discussed in Sec. \ref{srot}, contributed to the revival of the mirror concept. 

\begin{figure}[htp]
\centering\vspace*{-7mm}
\includegraphics[width=55mm,angle=-90]{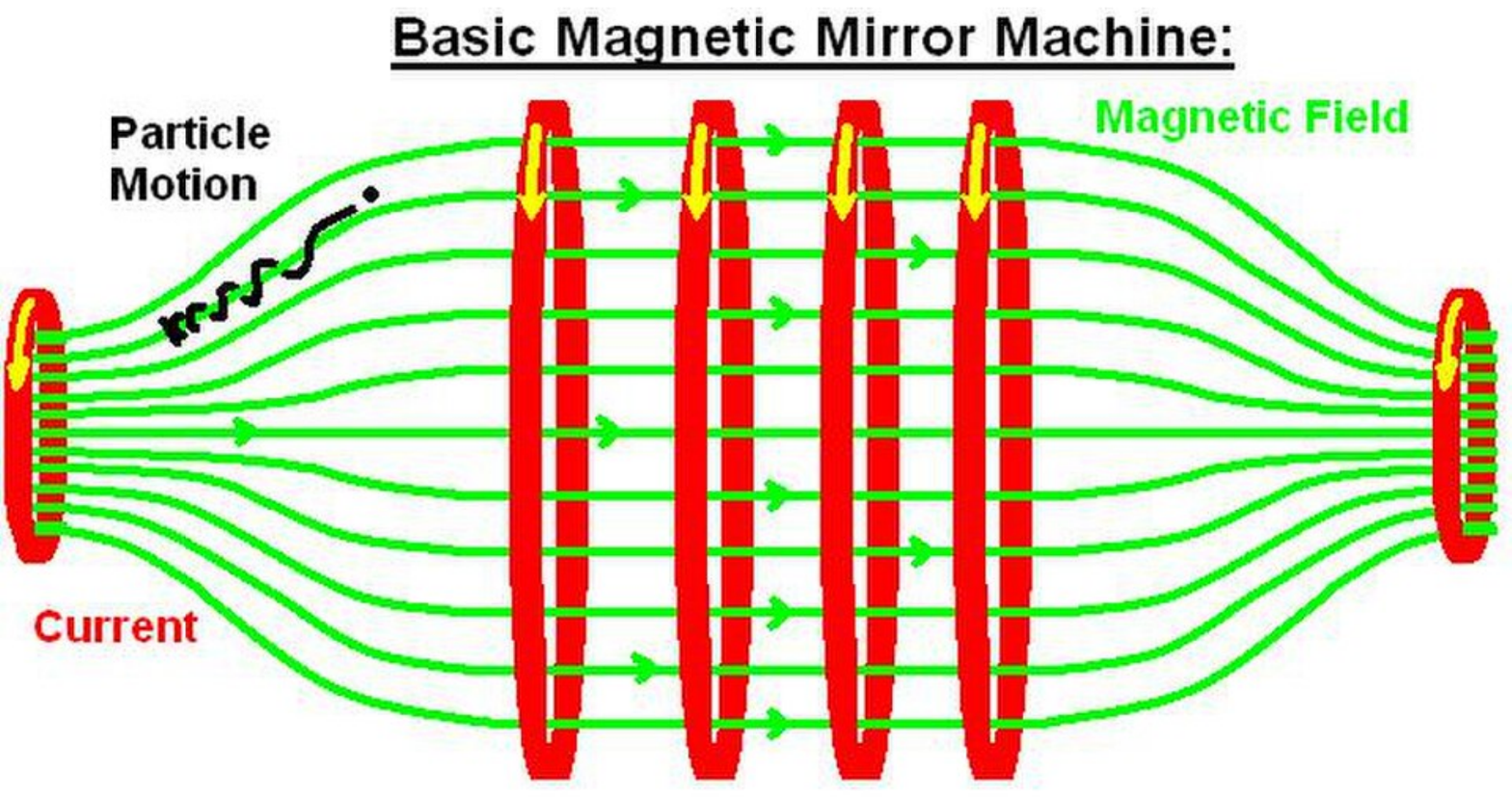}\vspace*{-7mm}
\caption{Basic mirror concept (Attribution: User: WikiHelper2134, via Wikimedia Commons).}
\label{mirror}
\end{figure}
The basic mirror concept is shown in Fig. \ref{mirror}. 
The confinement of charged particles in a mirror relies on 
the conservation of the magnetic moment $\mu$ and energy $\varepsilon$, where
\bea
\mu=&\frac {mv_\bot^2}{2B}~~~\hbox{and}~~~\varepsilon = \frac m2\lbs v_\bot^2 + v_\|^2\rbs 
\eea
with the subscripts $\bot$ and $\|$ denoting perpendicular and parallel to the magnetic field, respectively,
and $m$ being the mass \cite{post}. Denote the magnetic field strength at the mirror throats $z_m$ as $B_{max}$ and 
at an interior point $z_0$ as $B_0$. 
One can easily obtain the confinement condition for a charged particle in a mirror:
\bea
\frac {mv_\bot^2(z_m) }{2}=\frac {mv_\bot^2(z_0) }{2} \frac{B_{max}}{B_0} > \frac m2\lbs v_\bot^2(z_0) + v_\|^2(z_0)\rbs,
\label{conf}
\eea
i.e., $v_\|^2$ vanishes before reaching the throats. 
Here, the conservations of $\mu$ and $\varepsilon$ have been used. 
Using \eq{conf}, one obtains the confinement condition:
\bea
\frac{B_{max}}{B_0}>1 +\frac{v_\|^2(z_0)}{v_\bot^2(z_0)}.
\label{mfactor}
\eea 
The ratio of the maximum to minimum magnetic fields of a mirror, ${B_{max}}/{B_{min}}$, is referred to as
the magnetic mirror factor.

 \section{The  rotating mirror with detached electrodes}

\label{srot}

An important development of rotating mirror device is the invention of the detached 
electrodes to induce the electric field in a mirror. The electric field
together with the magnetic field makes the mirror plasma rotate. As explained later in
this section, a rotating mirror tends to stabilize the so-called interchange or flute modes,
which are the most concerning mirror instabilities. Therefore, fusion fuel plasma  
can be preheated in a rotating mirror before the pinch compressions, to be described
in Sec. \ref{spinch}, to be applied.

\begin{figure}[htp]
\centering\vspace*{-7mm}
\includegraphics[width=65mm,angle=-90]{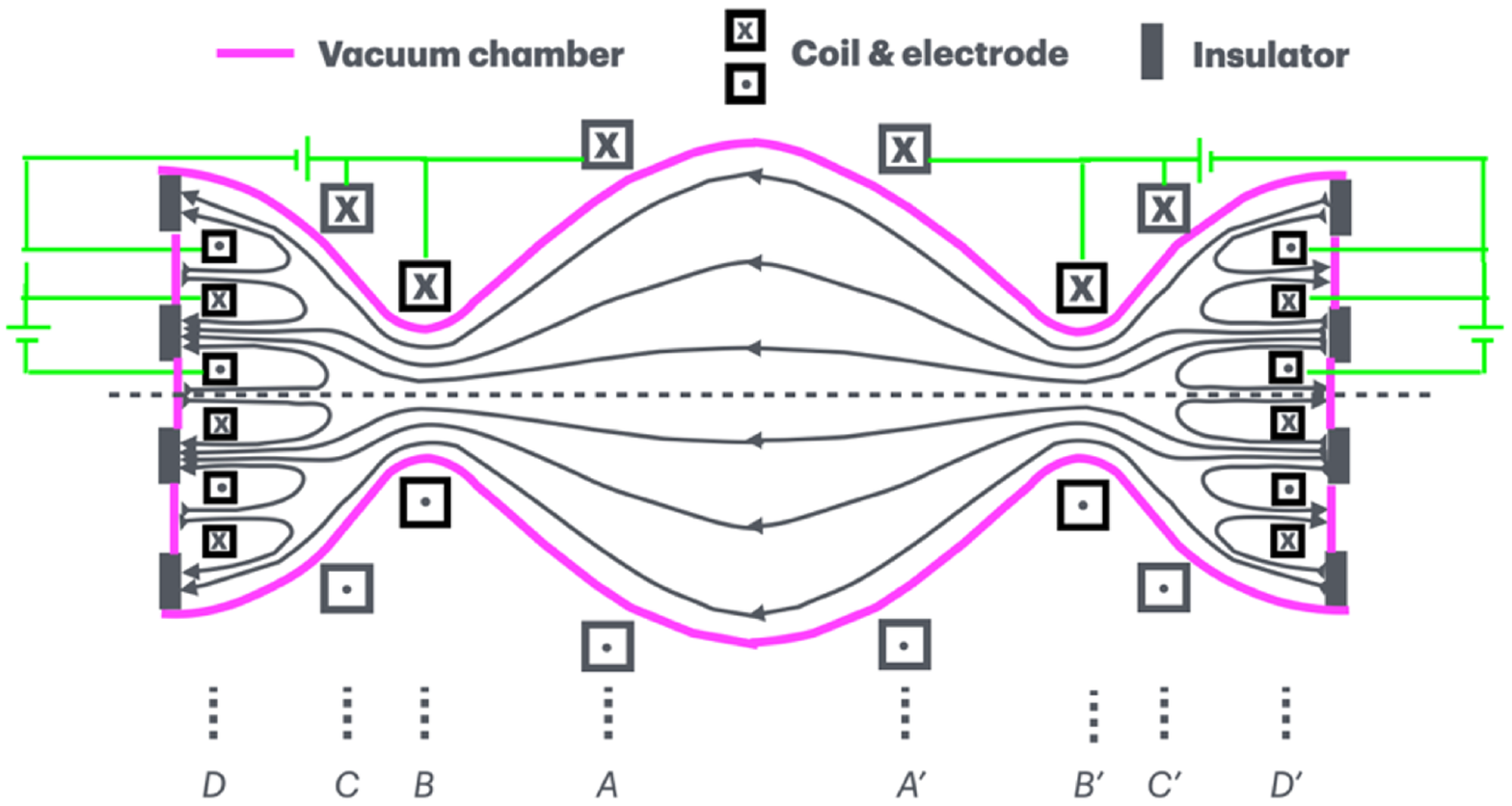}\vspace*{-7mm}
\caption{Mirror device with detached electrodes for driving the rotation. 
The system is azimuthally symmetric around the center axis, denoted by the horizontal dashed line
\cite{pt1}. }
\label{electrode}
\end{figure}

The invention of detached electrodes in a mirror in Ref. \cite{pt1} is explained in Fig.
\ref{electrode}. 
In this figure, the magnetic field lines are depicted in the black curves with arrows. The two coil sets on the left and right at positions “B” and “B’” are the main coils used to generate the magnetic mirror configuration with the current directions shown with “x” and “.”. More auxiliary coils in positions “A”, “A’”, “C”, and “C’” are introduced for controlling the shape. The coil/electrode sets in positions “D” and “D’” are used to apply the electric field bias, with the green lines describing the electric voltage applications, and to generate the separatrix magnetic field for detaching the electrodes from the magnetic field lines from the mirror core.  The insulators are introduced to avoid a short circuit due to the steel vacuum chamber. From Fig. \ref{electrode}, one can see
that the electrodes are shielded by the magnetic field from the bombardment of the hot 
charged particles from the mirror core. This is a more effective way to apply the electric field
in a mirror to generate the plasma rotation, as compared to the method using the insulators \cite{wham}. 

Let us now explain the physics mechanism for the rotation stabilization of the interchange or flute modes
in a mirror in Ref. \cite{stability}.  
First, we note that in the recent paper \cite{equilibrium} it is pointed out that the so-called
Hinton-Wong Boltzmann distribution function cannot be used for a mirror. This is because
there are end plates, mirror throats, and trapped particle effects. The usual solution
of the drift kinetic equation is no longer applicable in a mirror because the boundary conditions
need to be taken into account. Consequently, the azimuthal rotation frequency becomes non-constant
along the magnetic field lines.

\begin{figure}[h]\vspace*{-10mm}
\hspace*{-2mm}
\includegraphics[width=60mm,angle=-90]{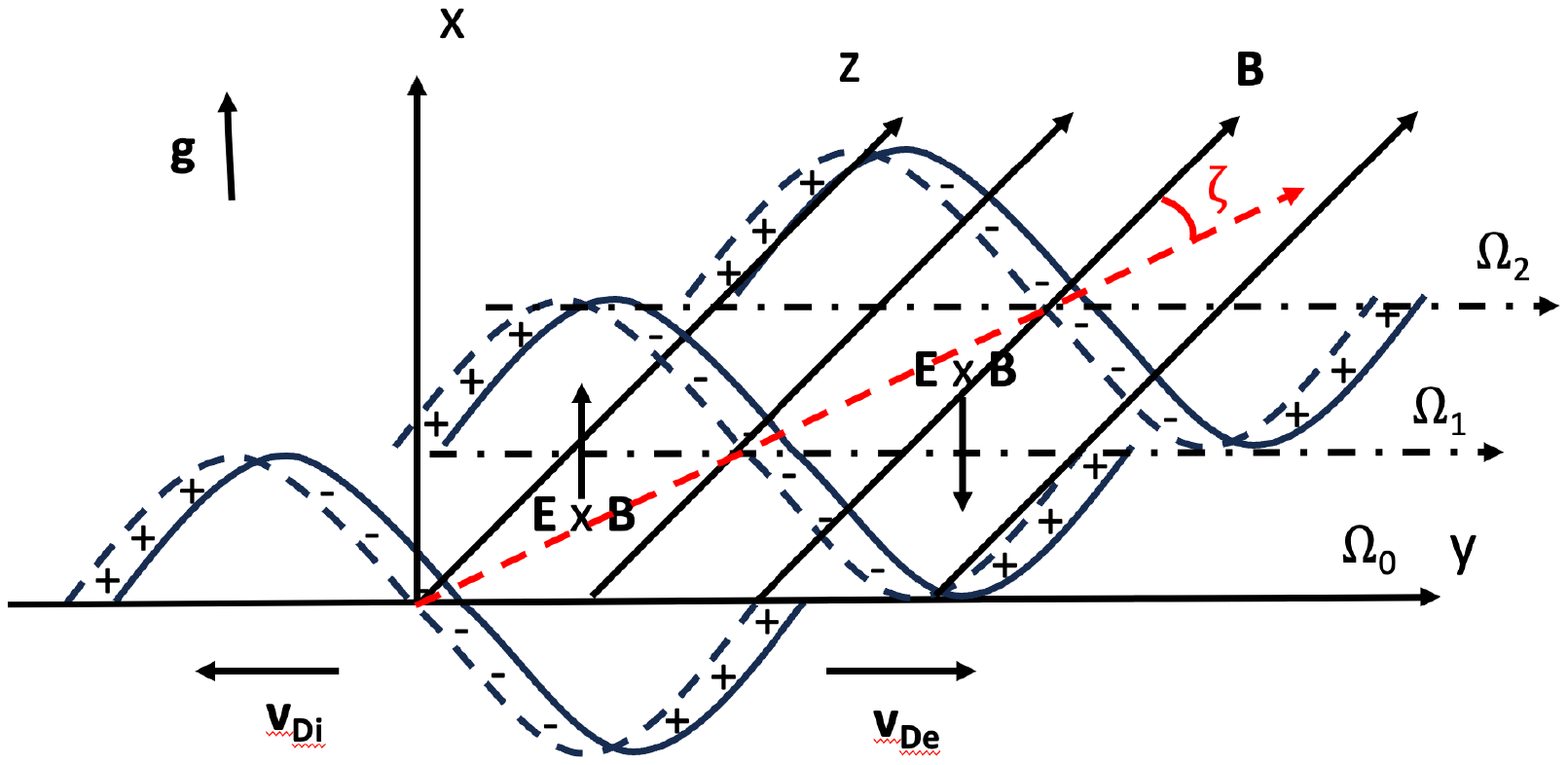}\vspace*{-9mm}
\caption{The short circuit effect on the interchange modes due to
the non-constant rotation along the magnetic field lines, i.e., when $\zeta\not =0$.
In this figure, the the gravitional force $\g$ is used to denote the magnetic curvature \cite{stability}.}
\label{fstability}
\end{figure}

Next, we note that there is a short-circuit effect on the interchange or flute modes due to
the non-constant rotation along the magnetic field lines. This can be explained in Fig.
\ref{fstability}. 
In this figure, $x$ corresponds to the radial direction, $y$ is the azimuthal direction, and
$z$ represents the longitudinal direction, with $(\e_x,\e_y,\e_z)$ representing the unit vectors for
each coordinate. The boldface is used to denote vectors. 
The magnetic field is in the $\e_z$ direction. 
The rotation frequency ${\bfg \Omega}$ is assumed to be in the $\e_y$ direction 
and is a function of $z$. For simplicity,
 the rotation frequency $\Omega(z)$ is assumed to be a linear function of $z$ 
 and is less than the sound wave frequency. 
 It is assumed that $\Omega_0=\Omega(0)
 < \Omega_1=\Omega(z_1)  < \Omega_2=\Omega(z_2) $. The ion and electron drift velocities
 are denoted by 
 $\v_{di}$ and $\v_{de}$, respectively, with
 \bea
 \v_d &=& \frac\b\Omega_c\btimes\lbs\mu\bnabla B +v_\|^2 \b\bcdot\bnabla\b\rbs,
 \label{vd}
\eea
where $\Omega_c=eB/m$ is the gyrofrequency.
 
When there are pressure gradient and unfavorable curvature, a fluid perturbation (slid curves)
can cause a charge separation (between the dashed and solid curves).
This is because the electron and ion drifts are in opposite directions as governed by \eq{vd}. 
The electric field induced by the separated charges
can further drive the initial interchange perturbation due to the $\E\times\B$ drift. 
This leads to an instability \cite{ros}.  
 If there is a parallel-sheared rotation, i.e., the rotation is non-constant
along the magnetic field lines, the wavefront is not perpendicular to the
magnetic field line, i.e., $\zeta\not=0$ in Fig. \ref{fstability}.
In this case, the separated changes caused by the drift motions 
will be neutralized due to the parallel mobility of electrons. This leads to the
stabilization of the interchange or flute modes.
These explain why the rotating mirror becomes stable to the interchange or
flute modes.

\section{The rotating mirror with all-directional pinch compressions}

\label{spinch}

In Sec. \ref{srot}, we have shown how the rotation can be driven in a mirror with
the detached electrodes and how the interchange or flute modes can be stabilized
in this system. The rotation stabilization of mirror confinement has been observed
experimentally with non-detached electrodes \cite{wham}. With the invention of
detached electrodes \cite{pt1}, one can expect better confinement to be achieved. 
However, this does not mean that one can simply continue to heat
the rotating mirror plasma to achieve nuclear fusion to get energy gain
as in the magnetic confinement fusion approach. There is a competition between the
heating power and the transport energy losses. Also, there are numerous plasma 
instabilities. One can only assure that the plasma rotation stabilizes the major instabilities,
such as the interchange or flute modes.
That is why the magnetic confinement fusion approach, including the rotating mirror direction,
remains challenging.  

Nevertheless, there is a promising approach by combining the steady-state and fast processes
based on the rotating mirror \cite{pt2}. The rotating mirror with detached electrodes
is used for steady-state preheating, and after the preheating, all-directional fast theta pinch compressions 
are applied. Although neither the steady-state heating of rotating mirror plasma 
nor the fast compression of the plasma on its own is foreseen to easily meet the Lawson criterion,
the proper combination of these two processes does. Before moving on to the details, let's make
a rough estimate here for the case with the theta pinch compression \cite{cpinch} applied on the
preheated fuel plasma based on the current experimental data. 

Let us briefly outline the parameters achieved in the current mirror experiments. The current simple axisymmetric mirror experiments can reach $1 – 10$ keV temperature with $0.5$ Tesla magnetic field.  The density can reach $10^{19}$ to $10^{20}$ 1/m$^3$. However, the confinement time is still in $1$ to a few $10$ milliseconds. The high-power neutral beam injection and radiofrequency heating are used to reach these parameters. We note that with the rotation introduced using the detached electrodes \cite{pt1}, one can expect that these parameters are achievable or can even be reached higher.
These parameters are lower than those achieved in modern tokamaks. However, they are sufficient as the preheated mirror plasmas for the theta-pinch compression to meet or exceed the Lawson criterion. 

\begin{figure}[htp]
\centering\vspace*{3mm}
\includegraphics[width=65mm,angle=-90]{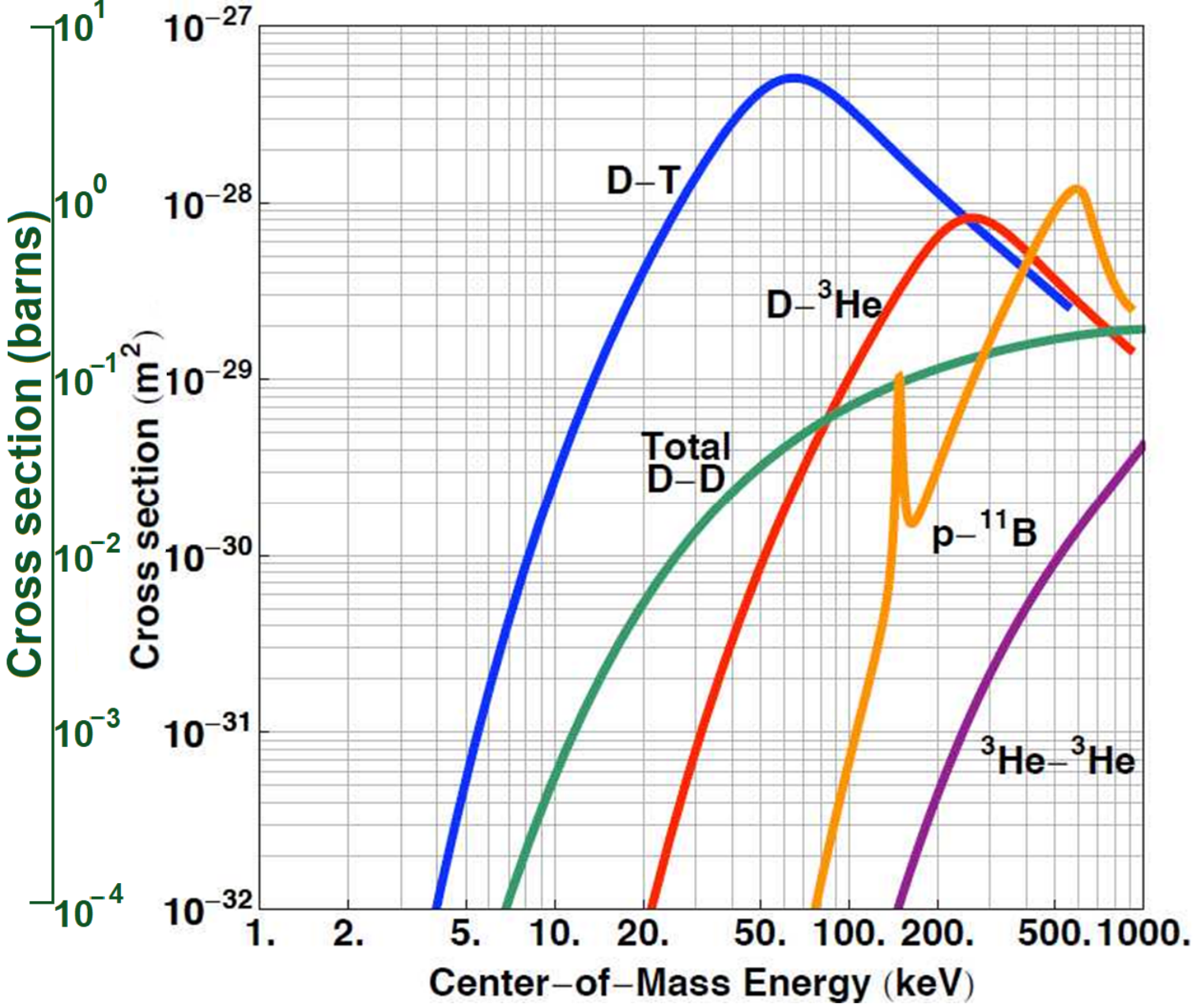}\vspace*{3mm}
\caption{Cross Section of DT, DD, D-He3, P-B11 and Helium3 nuclear fusion reaction. These rates are measured in cross section $(M^2)$ (Attribution: User: WikiHelper2221, via Wikimedia Commons).}
\label{fuse}
\end{figure}
The physics basis for the theta pinch compression is simply the conservation of magnetic moment, i.e.,  $\mu={mv}_{\perp }^{2}/2B$ is an invariant. This indicates that the square of the velocity perpendicular to the magnetic field, $v_{\perp }^{2}\sim T_\bot$, is proportional to the magnetic field strength $B$. When the magnetic field increases in the pinch process, the ion kinetic energy increases proportionally. The principle is solid and well-proven. With the current superconductor technology, the magnetic field can be raised to $20$ Tesla or more. Because of the conservation of the magnetic moment, one can expect the perpendicular 
temperature $T_\bot$ is raised to $40$ to $400$ keV. 
Using the fusion cross sections diagram in Fig. \ref{fuse}.
one can see that this is far more than enough for D-T fusion, which peaks at about $70$ keV
while $10-20$ keV is usually good enough. 
One can even consider the p-11B fusion, which peaks at around $500$ keV.
In the p-11B fusion case, there is no neutron byproduct, which further reduces the technological difficulty by avoiding the neutron treatment. 

\begin{figure}[htp]
\centering\vspace*{-14mm}
\includegraphics[width=65mm,angle=-90]{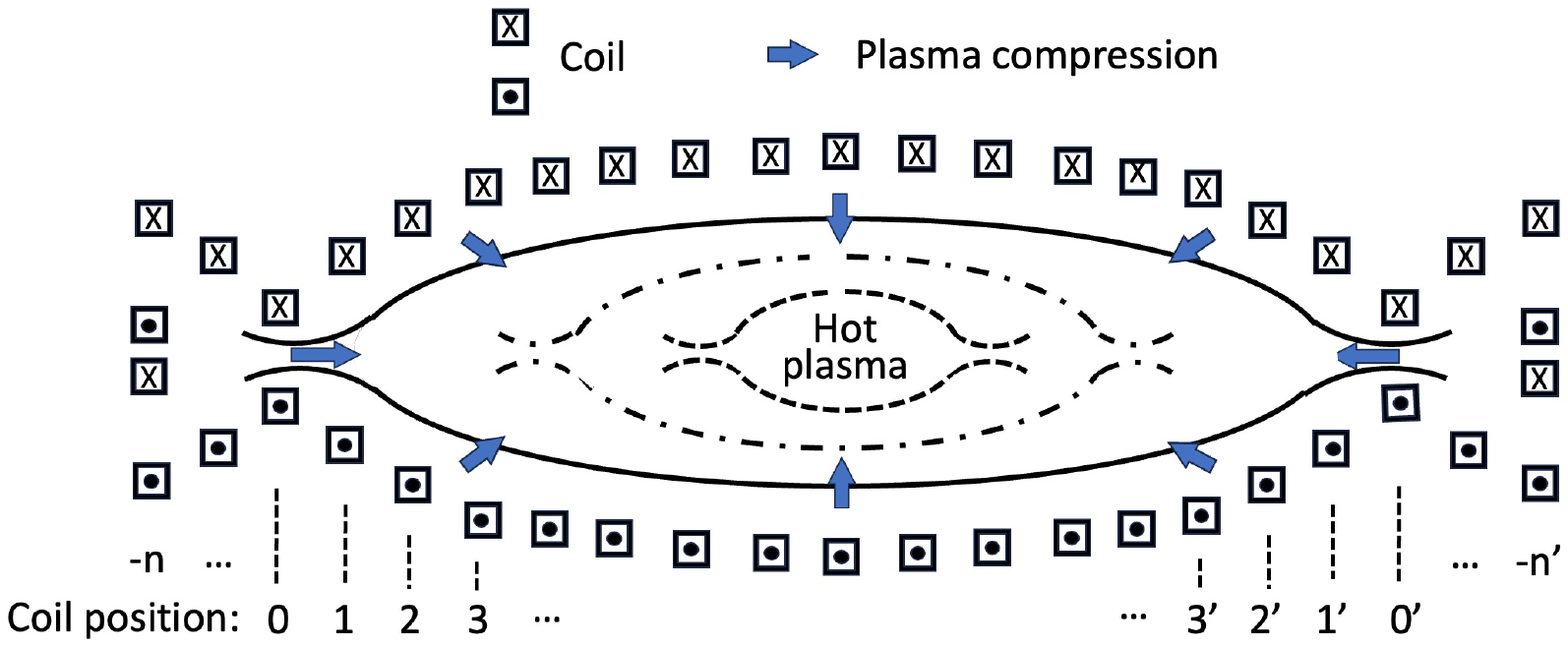}\vspace*{-12mm}
\caption{Rotating mirror device with both radial and longitudinal pinch compressions. It is azimuthally symmetric around the horizontal center axis \cite{pt2}.}
\label{comp}
\end{figure}
The preheating in the rotating mirror with detached electrodes has been described in
Sec. \ref{srot}. Let us now describe how the all-directional pinch compressions
are applied to the preheated fuel plasma.
As shown in Fig. \ref{comp}, after the mirror plasma is preheated, the two-directional compressions in both the radial and longitudinal directions are implemented. Since the system is azimuthally symmetric, 
two-directional compressions are the all-directional compressions.
The all-directional compressions are realized by raising the coil currents sequentially with proper amounts.
 The original magnetic-mirror-confined plasma, which is preheated, is compressed in both the radial and longitudinal directions. The volume of mirror-confined plasma is reduced step by step until a targeted shape is reached. The sequential process appears to be like the two pinch guns moving toward each other. Three compression sequences are plotted in Fig. \ref{comp}: the original preheated one denoted by the solid curves, the later one denoted by the dot-dashed curves, and the further later one by the dashed curves. Multiple coils are used for this purpose. The current peaks move sequentially and inwardly from both the horizontal sides from positions “0” and “0’”, to positions “1” and “1’”, and further inward to positions “2” and “2’”, and so on. This makes the mirror throats shrink inwardly from both the left and right sides to the targeted positions. The individual coils are biased electrically with respect to the detached electrodes at the mirror ends at positions “-n” and “-n’” as described in Sec. \ref{srot} \cite{pt1} to induce the plasma rotation.

To justify the engineering feasibility of the longitudinal pinch compression, let us examine 
the change of the magnetic mirror factor after the pinch. 
Define the pinch factor as
\bea
P =\lbs  \frac{B_{0,new}}{B_{0,orig}}\rbs_{pinch},
\eea
where the subscripts "new" and "orig" have been used to denote the values after and before the
pinch process.
Due to the conservation of  the magnetic moment,  the perpendicular kinetic energy is raised
while the parallel one remains unchanged, i.e.,
\bea
v_{\bot,new}^2 = P v_{\bot,orig}^2~~~\hbox{and}~~~ v_{\|,new}^2 =  v_{\|,orig}^2.
\label{pp}
\eea
According to \eq{mfactor}, the mirror factor is given by
\bean
\lbs \frac{B_{max}}{B_0}\rbs_{after~pinch}&=& 1+\frac{ v_{\|,rew}^2}{v_{\bot,new}^2} 
\nn
& =& 1+ \frac{ v_{\|,orig}^2}{P v_{\bot,orig}^2 } .
\eean
This leads to
\bea
\frac {\lbs \frac{B_{max}}{B_0}\rbs_{after~pinch}-1}
{\lbs \frac{ B_{max}}{B_0}\rbs_{orig}-1} &=& \frac1P=  \lbs \frac{B_{0,orig}}{B_{0,new}}\rbs_{pinch}.
\label{pinch}
\eea
Here,  Eqs. \eqn{mfactor} and \eqn{pp} have been used. Equation \eqn{pinch} indicates that
the required mirror factor for confining the preheated charged particles before the pinch 
is reduced dramatically. Suppose the original mirror factor is the same as the pinch factor. 
Then, the required new mirror factor is less than $2$. This makes the longitudinal
compression easy to facilitate. The axial losses have been a concern to the radial theta pinch
compression. The longitudinal compression also provides axial plugs to constrain the 
axial losses,

How to determine the coil currents in each pinch stage has been described in Ref. \cite{bookmhd}.
Both the vacuum region between the plasma and the wall/coils and the vacuum region
external to the wall/coils are governed by the two-dimensional Laplace equation
\bea
\bnabla^2\varphi=0,
\label{phi}
\eea
where $\B=-\bnabla \varphi$. On the plasma-vacuum interface, the normal and tangential
components of the magnetic field are required to be continuous. The two-dimensional Laplace
equation, \eq{phi}, can be integrated from the plasma-vacuum interface toward the wall/coils position.
In this way, one obtains the vacuum solution inside the wall/coils. The vacuum solution outside
the wall/coils can be obtained by requiring $\varphi$ to vanish (or to be small) at infinity. 
 The normal 
component of the magnetic field is required to be continuous across the wall/coils.
The normalization constant can be used to satisfy it. The 
jump of the tangential component of the magnetic field determines the coil currents in the
inner and outer solution matching. 

One can understand that all-directional compressions are more effective
than the one-directional compression as in the theta pinch case.
However, there is a deeper physics reason to consider the two-directional
compressions. In the preheating process, the magnetic curvature should be reduced for stability. Therefore, a long and thin mirror configuration is favored \cite{gdt0,gdt}. When plasma is pinch-compressed, the precession drift plays an important role in stability. Therefore, it is favored to compress the plasma longitudinally. 
One can draw a comparison with the so-called ELMO bumpy torus (EBT) \cite{ebt}.
In the EBT case,  electrons are heated to achieve the precession drift stabilization. 
The hot electron temperature is about $400$ keV. 
From the drift velocity expression in \eq{vd}
one can see that the drift velocity is proportional to the mass. The proton-to-electron mass ratio is about 1836. 
To have the same precession frequency of electrons in
EBT, $1$ keV ion temperature is enough, not to mention the perpendicular temperature is much bigger after pinch compressions if preheating is implemented, which leads to a larger magnetic curvature. 
This should be related to the stability of the early Scylla IV-P experiments \cite{scy}.
The plasma has not been preheated in the Scylla experiments. Only the so-called wobble
instabilities were observed, which are nonlinearly stable and can be avoided,

The other interesting feature is that the all-directional pinch compressions can definitely 
result in nuclear fusion reactions. Consequently, high-energy alpha particles are generated. 
The precession drift of high-energy alpha particles can also
help stabilize the interchange or flute modes.

\begin{figure}[htp]
\centering\vspace*{-14mm}
\includegraphics[width=65mm,angle=-90]{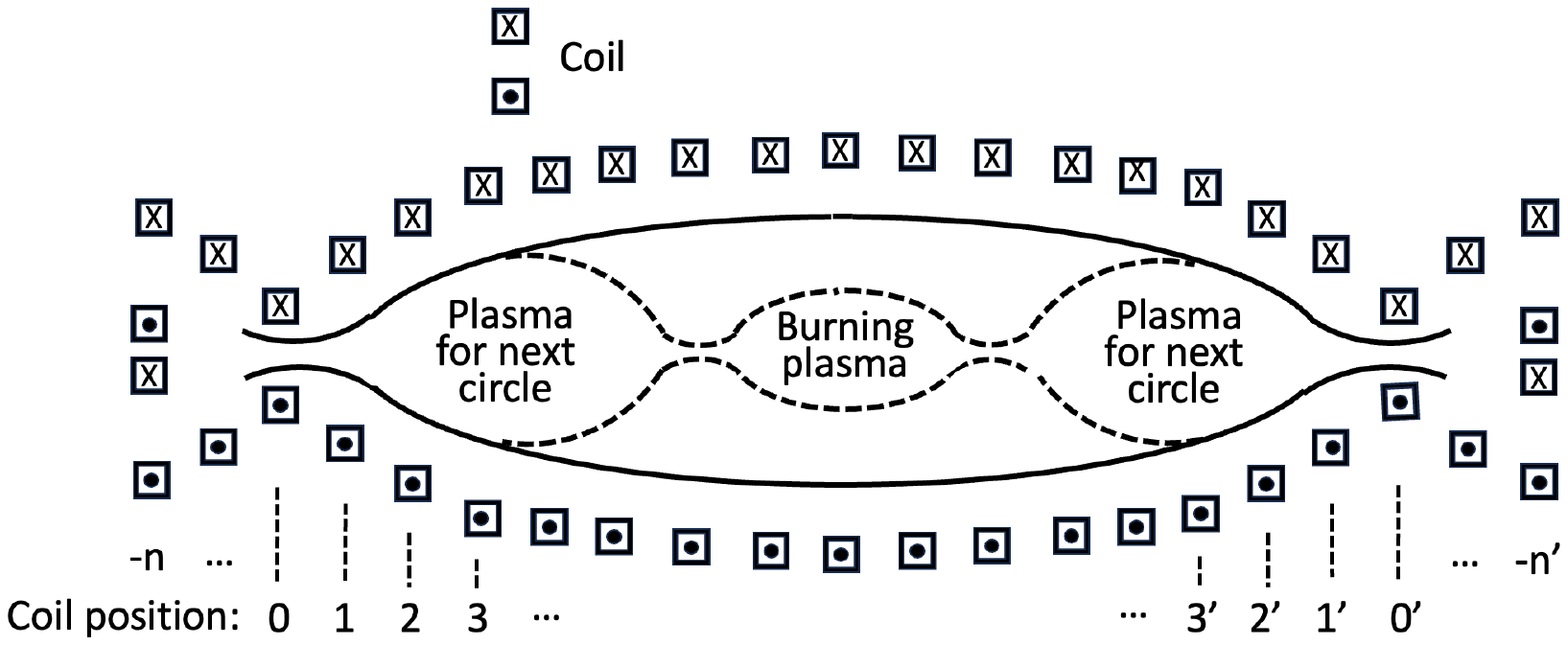}\vspace*{-13mm}
\caption{The final stage after the radial and longitudinal pinch compressions
from the processes described in Fig. \ref{comp}.
 It is azimuthally symmetric around the horizontal center axis \cite{pt2}.}
\label{burning}
\end{figure}
Note that in the current approach, the currents are maintained after compressions.
With the superconducting coils, this is feasible. This is another key difference from
the early Scylla IV-P experiments, besides the preheating, the rotating plasma, 
and two-directional compression. Figure \ref{burning} shows the final target
confinement configuration. Fusion occurs in the burning plasma core. The remaining
two sections on the left and right are the leftover plasmas to be heated for the next circle.
Note that because of the conservation of the magnetic moment, the plasma
perpendicular temperature increases with $B$, but the plasma beta $nT/B^2$ varies
along $B^{-2}$. Therefore, the plasma beta reduces after the compression.  
The force balance in the new stage requires the density to increase.

Before further exploring the features of the current approach, let us now estimate Lawson's criterion.
For a deuterium-tritium (D-T) reaction,  Lawson's criterion expressed in 
the fusion triple product is round \cite{lawson}
\bea
n\cdot T\cdot \tau > 5\times 10^{21} ~M^{-3}\cdot keV\cdot s.
\label{lawson}
\eea
where $n$ is density, $T$ is temperature, and $\tau$ is the energy confinement time.
Using the parameters achievable in the Gas Dynamic Trap (GDT) \cite{gdt0,gdt}:
the magnetic field $B\sim 0.35-0.5$ Tesla in the central chamber, the density
$n \sim 1 \times 10²^{20}$ M$^{-3}$, electron temperature $T_e\sim1$ keV, 
ion temperature $T_i\sim 1 - 10$,  keV, and energy confinement time  
 $\tau\sim 0.0005-0.006$ s. These parameter ranges for density, temperature, and
 confinement time also appear in other mirror devices, for example, in WHAM \cite{wham}.  
 Using the ion temperature, GDT achieves a typical fusion triple product of 
 \bea
 n\cdot T_i\cdot \tau  \approx 6\times 10^{17} ~ M^{-3}\cdot ke\cdot s.
 \label{gdt-lawson}
 \eea
 Here, 1 keV ion temperature has been used. The reason for using the ion temperature
 is explained as follows. 
 Lawson's analysis is based on the rate of fusion and loss of energy in a thermalized plasma. 
For magnetic confinement fusion based on the steady-state heating process, 
the electron radiation losses are larger than the ion ones. Therefore, the electron temperature
is used to estimate the Lawson criterion. 
The fast-process approaches, instead, directly accelerate or compress the ions to the required energies
well before the thermalization can actually occur,
The colliding beam fusion reactor Migma \cite{migma} or the inertial confinement
fusion are examples. In this case, using the ion temperature is more relevant. 
The current all-directional pinch-compressed fusion belongs to this case. This is another
advantage of succeeding the preheating of fuel plasma with a fast pinch compression.
Fast compression limits the time for thermalization between ions and electrons
and therefore minimizes the impact of electron bremsstrahlung radiation loss on ions.
 
\begin{figure}[htp]
\centering\vspace*{3mm}\hspace*{-5mm}
\includegraphics[width=75mm,angle=-90]{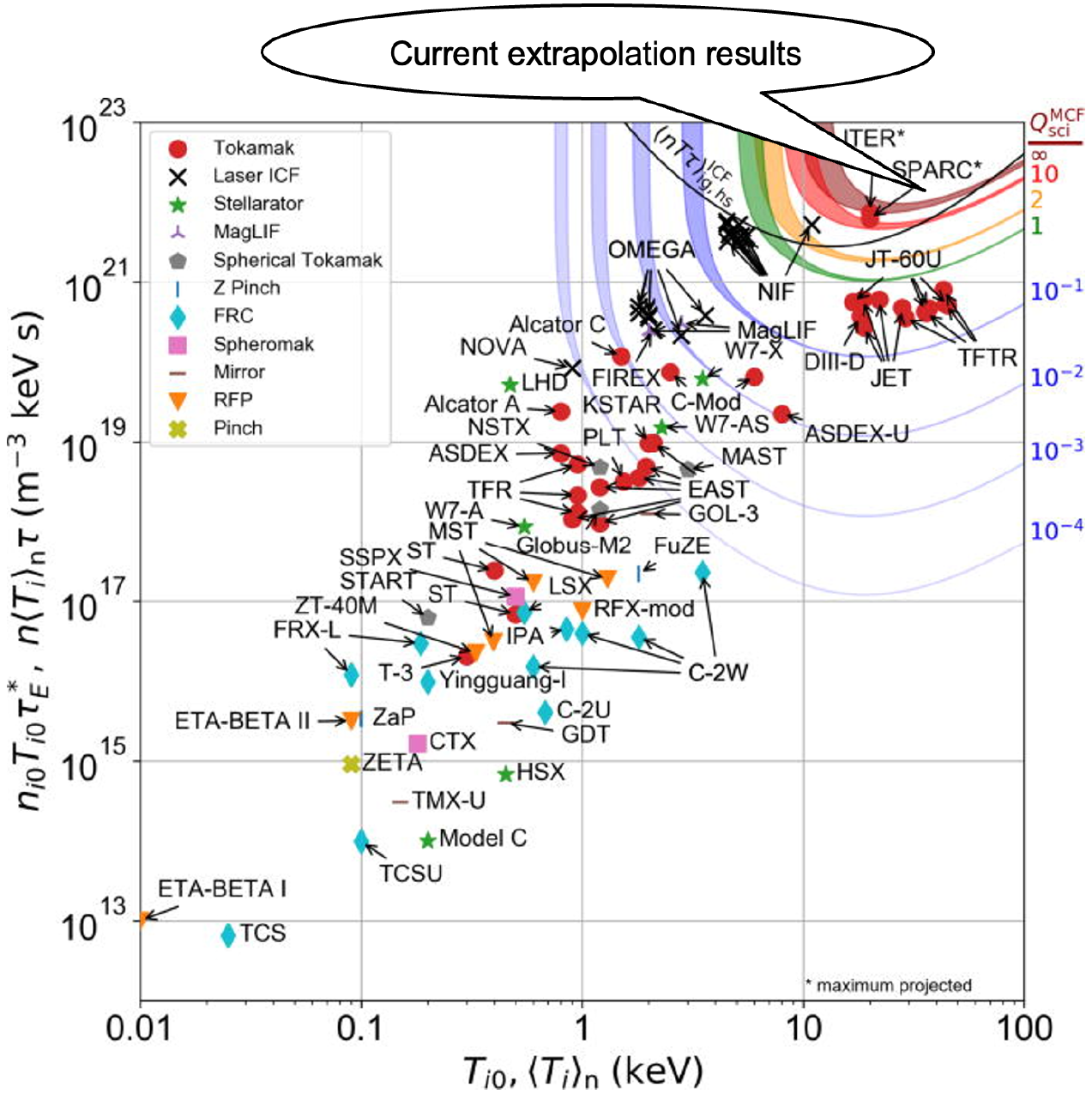}\vspace*{3mm}
\caption{The triple product and fusion gain factor $Q$ for various devices, reproduced from 
Ref. \cite{ctriple}. The extrapolation results of the current device is added, which
indicate the potential to reach or exceed the Lawson criterion.}
\label{triple}
\end{figure}

Suppose the magnetic field is raised to $20$ Tesla. Due to the conservation of 
 the magnetic moment, the ion and electron perpendicular 
 temperatures will be raised by a factor $40$.  Because of force balance,
 the plasma density also scales with the magnetic field $B$.
 The density is also increased by $40$ due to the radial pinch.
 One can assume the longitudinal density compression factor to be $10$, noting that
 GDT is 7 meters long. Therefore, the total amplification factor is about $16,000$.
 Multiplying this factor to the preheated value in \eq{gdt-lawson}, one can see that
the triple product in \eq{lawson} is exceeded (see Fig. \ref{triple}).
 
Note here that the GDT parameters given above are achieved based on the
simple mirror concept without introducing the plasma rotation, especially
with the detached electrodes. The WHAM's experiments confirm that the rotation
can improve the confinement \cite{wham}. Also, after the two-directional pinch compression,
the procession drift stabilization can enhance the confinement time.
This is not reported in the Scylla IV-P experiments since the coil currents
are not maintained after the pinch process. With the superconducting
coils, the final stage can be maintained with constant coil currents. 
The longitudinal compression also prevents the axial losses, 
These further show that the current method is promising. 

Let us return to discuss the additional features of the current method. 
It is intriguing to compare with the tokamak Ohmic heating. In tokamaks, plasma is heated through the inductive current. In mirrors, the magnetic moment invariant leads to the ions and electrons being heated up directly by the perpendicular temperature. With the same power, one can expect that the mirror theta pinch process is much more effective. Furthermore, the tokamak Ohmic current is transient. Without a drive,
the Ohmic current will die out. In the mirror, however, the currents flow in the superconducting coils. 
The currents can be maintained. Also, there are no disruptions of tearing mode instability issues.

It is intriguing to discuss some engineering features. The two-directional compressions
make the burning plasma shrink as shown in Fig. \ref{burning}. This avoids 
direct contact of hot plasma with the external solid-material components, 
such as the wall, coils, etc.  The open-end mirror configuration
is easier to couple to the MHD generators for direct
energy conversion to electric power as implemented recently in WHAM.
It is also easier to remove fusion ashes.

\section{Conclusions and discussion}

\label{scon}

This paper describes the rotating mirror device with detached electrodes and all-directional pinch compressions as a shortcut to controlling nuclear fusion as an energy source. It is based on the two provisional patents filed recently by the University of Texas at Austin. The device combines the steady-state and fast processes in the two main streams of controlled nuclear fusion research: magnetic confinement fusion and inertial confinement fusion. 
The fuel plasma is preheated in a steady-state process in a rotating mirror with detached electrodes
and then all-directional pinch compressions are applied as the fast process. 
Preheating and longitudinal compression, in addition to the radial compression, significantly boost the nuclear fusion rate. Fast compression after the preheating limits the time for thermalization between ions and electrons and, therefore, minimizes the impact of electron 
bremsstrahlung radiation loss on ions.
Based on the existing experimental results,  the current method can be extrapolated to have the potential to reach or exceed the Lawson criterion for the first demonstration of the feasibility of peaceful usage of nuclear fusion energy (see Fig. \ref{triple}).

 It is interesting to compare with various existing concepts. The ITER tokamak has a plasma major radius of 6.2 meters. The circumference of its plasma torus is about 39 meters. Constructing a 39-meter-long mirror is much simpler than constructing a tokamak that requires inductive current. Compressing a 39-meter-long mirror longitudinally to 1-4 meters is reasonably achievable. Compared to the conventional linear or toroidal theta pinch or Z pinch, the current method generates an order of magnitude additional compression factor through the longitudinal compression. Unlike the concepts of shooting two preheated plasmoids toward each other at a short distance [9-11], the current method holds the preheated plasma in the central core of a long mirror and then pinches it radially and longitudinally. The current method again adds a large longitudinal compression factor. Combining steady-state confinement for preheating with fast pinch compression is a shortcut to satisfying the Lawson criterion, while the longitudinal compression factor makes it possible. The conventional inertial confinement fusion uses laser or particle beams to compress a tiny pellet of fuel that holds a few milligrams of frozen hydrogen isotopes. The current method, however, uses a magnetic pinch process to compress large amounts of preheated plasma.

Controlling nuclear fusion is so challenging that for decades, people have been asking: Are we closer to infinite clean energy? Significant progress has been made over the decades in this field in terms of 
achieving the triple product as required by the Lawson criterion. The leading positions 
 for magnetic confinement fusion and inertial confinement fusion are
 ITER (as well as SPARC) and NIF, respectively.
They are reaching or slightly above the breakeven 
($Q\sim 1-10$, see Fig. \ref{triple}). But still considerable efforts are needed to 
satisfy the Lawson criterion. Science may not only be
measured solely by the final successes. Step by step, tremendous efforts in this field 
also contribute to a much better understanding of the underlying physics,
the plasma and controlled nuclear fusion physics. Mirror and pinch are parts of these efforts.
The current work shows that, based on the existing progress, it is worthwhile to examine them
and combine individual advantages to find a shortcut. The Creator's Wisdom, beauty and simplicity,
is always the ultimate guidance. The limitation is only in our understanding. Looking forward, even with
the current method, there could still be unforeseen challenges or difficulties. Let me use 
Psalm 123:2 again as Ref. \cite{adv} to conclude this paper:

As the eyes of slaves look to the hand of their master, as the eyes of a maid look to
the hand of her mistress, so our eyes look to the LORD our God, till He shows us
His mercy.

He has the clue…  

\vspace*{3mm}

\noindent {\sf Acknowledgements:} 
The author would like to acknowledge Professor R. D. Hazeltine, 
Dr. M. T. Kotschenreuther, and other colleagues for the helpful discussions.
This research is supported by the U. S. Department of Energy, Office of Fusion Energy Science
under Grant No. DE-FG02-04ER54742.


\end{document}